\documentclass[prl,twocolumn,superscriptaddress]{revtex4}
\usepackage{amsmath}
\usepackage{bm}
\usepackage{graphicx}
\usepackage{color}
\usepackage{float}
\usepackage{graphicx}

\begin{document}

\title{Discontinuous fluidisation transition in assemblies of
actively-deforming particles: A new paradigm for collective motion
in dense active materials}

\author{Elsen Tjhung}

\affiliation{Laboratoire Charles Coulomb, UMR 5221, CNRS and 
Universit\'e Montpellier, Montpellier, France}

\affiliation{Department of Applied Mathematics and Theoretical Physics, University of Cambridge, Cambridge CB3 0WA, UK}

\author{Ludovic Berthier}

\affiliation{Laboratoire Charles Coulomb, UMR 5221, CNRS and 
Universit\'e Montpellier, Montpellier, France}

\date{\today}


% 05.40.-a    Fluctuation phenomena, random processes, noise, and Brownian motion
% 05.65.+b 	Self-organized systems
% 47.57.E- 	Suspensions

\begin{abstract}
Tracking experiments in dense biological tissues reveal a diversity of sources for local energy injection at the cell scale. 
The effect of cell motility has been largely studied, but much less is known about the effect of the observed volume fluctuations of individual cells. 
We devise a simple microscopic model of `actively-deforming' particles where local fluctuations of the particle size constitute a unique source of motion.
We demonstrate that collective motion can emerge under the sole influence of such active volume fluctuations.
We interpret the onset of diffusive motion as a nonequilibrium first-order phase transition, 
which arises at a well-defined amplitude of self-deformation. 
This behaviour contrasts with the glassy dynamics produced by self-propulsion, 
but resembles the mechanical response of soft solids under mechanical deformation. 
It thus constitutes the first example of active yielding transition. 
\end{abstract} 

\maketitle

Active matter represents a class of nonequilibrium systems that is currently under intense scrutiny~\cite{ramaswamy-review,vicsek-review}.
In contrast to externally driven systems (such as sheared materials), 
active matter is driven out of equilibrium at the scale of its microscopic constituents. 
Well-studied examples include biological tissues~\cite{chavrier-tissues}, 
bacterial suspensions~\cite{weibel} and 
active granular and colloidal 
particles~\cite{ramaswamy-vibrated-rods,olivier,cecile,clemens}.

Epithelial tissues constitute a biologically relevant active system composed of densely packed eukaryotic cells~\cite{shraiman,trepat-glassy-tissue,mehes,chavrier-tissues,silberzan,rorth,basan}.
Such tissues display a surprisingly fast and collective dynamics, which would not take place under equilibrium conditions~\cite{mehes}. 
This dynamics has been ascribed to at least three distinct active processes~\cite{rorth}:
(i) self-propulsion through cell motility such as crawling~\cite{theriot},
(ii) self-deformation through protrusion and contraction~\cite{zehnder,la-porta,yamamoto}, and
(iii) cell division and apoptosis~\cite{basan}.
The vertex model for tissues~\cite{shraiman,julicher,manning} includes the first two of these active processes and 
predicts a continuous static transition from an arrested to a flowing state~\cite{manning2}.
Another theoretical line of research is based on self-propelled particles~\cite{vicsek} 
which display at high density a nonequilibrium glass transition~\cite{berthier-nat-phys,marchetti-active-jamming} 
accompanied by a continuous increase of space and time correlations which diverge on approaching the arrested phase~\cite{RanNi,berthier-SP,szamel}.
However, typical correlation lengthscales in tissues do not seem to diverge~\cite{trepat-glassy-tissue,silberzan,garcia,vedula}.

To disentangle the dynamic consequences of the various sources of activity in tissues at large scale, 
we suggest to decompose the original complex problem into simpler ones, 
and to study particle-based models which only include a specific source of activity. 
This strategy was followed earlier for self-propulsion, 
but experiments are instead often modelled by complex models with many competing processes~\cite{sepulveda,garcia,yamamoto}. 
We argue that it is relevant to introduce a simplified model 
to analyse the effect of active particle deformation in a dense assembly of non-propelled soft objects.  
Specifically, we model a dense system that is driven out of equilibrium locally 
through `self-deformation' rather than self-propulsion. 
We study soft particles that actively change their size, 
while energy is being dissipated through viscous damping.
As a starting point, we consider the simplest form of self-deformation, 
in which the diameter of each spherical particle oscillates at very low frequency, 
in a way that is directly inspired by experimental observations in real tissues~\cite{zehnder}. 
The interest of such modelling is that activity is thus controlled by a unique adimensional parameter, $a$, 
which quantifies the relative change of the particle diameter within a deformation period. Our aim is not to propose a realistic model of a tissue, 
but rather to answer a more fundamental physical question regarding the role
of active volume fluctuations in tissue dynamics.  
Despite its relevance, such a class of models has, to our knowledge, 
not been analysed before in a statistical mechanics context.

Our main result is the existence of a discontinuous non-equilibrium phase transition 
from an arrested disordered solid to a flowing fluid state at some critical activity, $a_c$, 
with no diverging timescales or lengthscales.
In particular, we observe a modest increase of one order of magnitude 
in the relaxation times of the fluid before the system gets 
discontinuously trapped in an arrested phase at $a = a_c$. 
Our system also shows a strong hysteresis as seen in equilibrium first-order phase transitions.
This scenario for the fluidisation transition differs markedly from that of self-propelled particles 
in which a dramatic continuous slowing down is observed~\cite{berthier-nat-phys,RanNi,berthier-SP,szamel}. 
We propose that the correct analogy for our observations is not with a glass \cite{berthier-nat-phys} or 
jamming~\cite{marchetti-active-jamming} transition, 
but rather with the yielding transition of amorphous solids~\cite{bonn}, 
which start flowing irreversibly when mechanically perturbed beyond a force threshold, the yield stress. 
Therefore, actively-deforming particles undergo an `active yielding transition', which represents
a novel paradigm for the collective motion of active materials.

We consider a dense suspension of $N$ soft circular particles at 
zero temperature in a two-dimensional square box of linear size $L$ with
periodic boundary conditions.
The interaction between the particles is modelled by a short-ranged 
repulsive harmonic potential, similar to jammed foams~\cite{durian}:
$V(r_{ij})=\frac{\epsilon}{2} \left( 1- r_{ij}/ \sigma_{ij} 
\right)^2 H(\sigma_{ij}-r_{ij})$,
where $r_{ij}=|\vec r_i - \vec r_j|$,  $\sigma_{ij}=(\sigma_i + \sigma_j)/2$, 
with $\sigma_i$ and $\vec r_i$ the diameter and position of particle $i$, 
respectively.
The energy scale of the repulsive force is set by $\epsilon$,
and $H(x)$ is the heaviside function, defined such that $H(x \ge 0)=1$.
In the overdamped limit, the dynamics of each particle is described 
by a Langevin equation:
\begin{equation}
\xi\frac{d\vec r_i}{dt}=-\sum_{j\neq i}\frac{\partial V(r_{ij})}{\partial 
\vec r_j},
\label{eq:motion}
\end{equation} 
where $\xi$ is a friction coefficient.
The dissipation timescale is $\tau_0= \xi \sigma_0^2/\epsilon$, 
where $\sigma_0$ sets the particle diameter (see below).
Physically, $\tau_0$ is the typical timescale for a system 
described by Eq.~(\ref{eq:motion}) to come at rest without 
forcing.

\begin{figure}
\begin{centering}
\includegraphics[width=8.5cm]{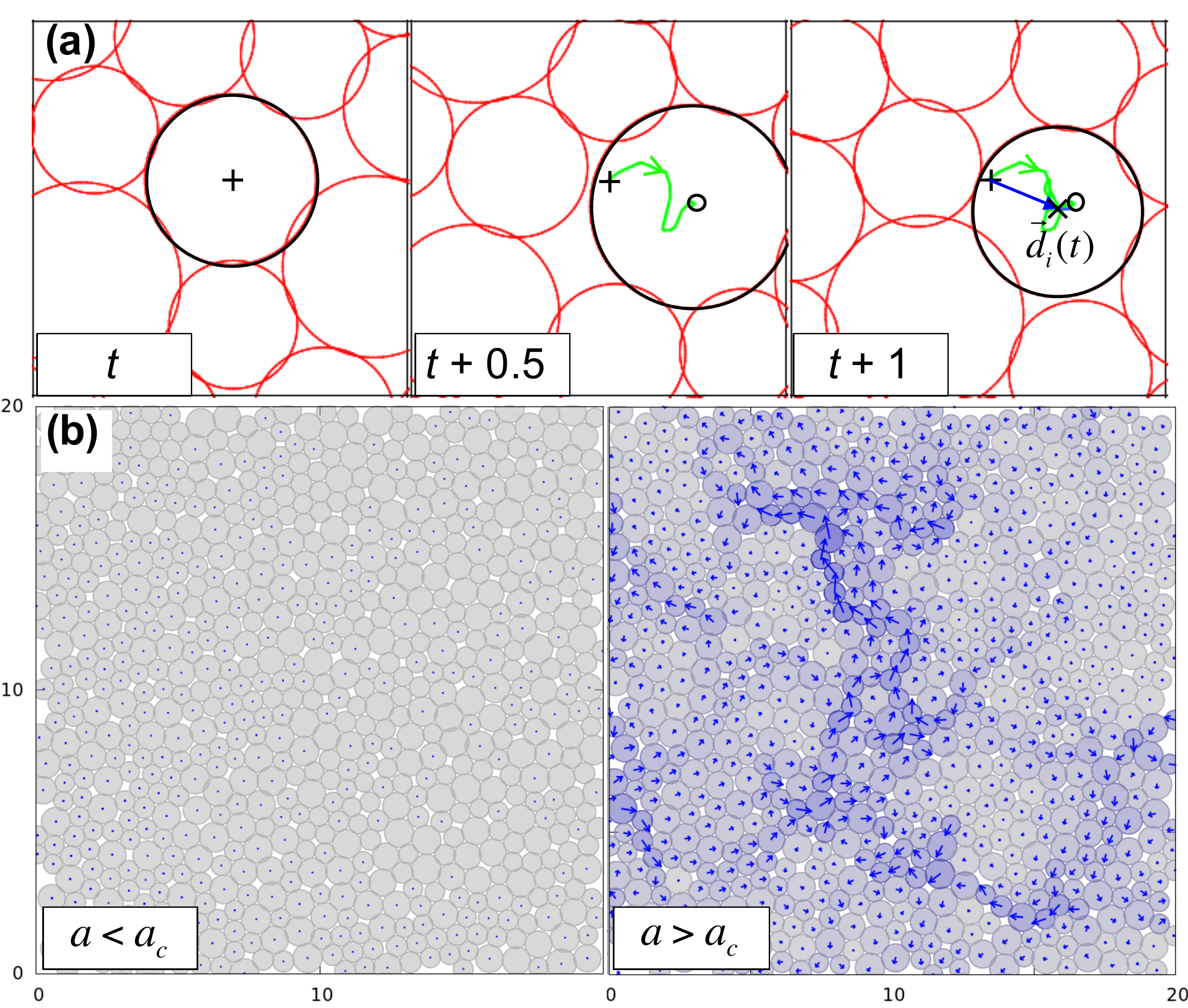}
\par\end{centering}
\protect\caption{(a) Snapshots of the system over one cycle of 
active deformation. 
The green curve is the trajectory of the highlighted particle 
during one cycle. The blue arrow represents its 
displacement after one cycle $\vec d_i(t)= \vec r_i(t+1)-\vec r_i(t)$.
(b) One-cycle displacement map $\vec d_i(t)$ in steady state for the 
disordered solid phase ($a=0.047<a_c \approx 0.049$, left) and in the 
fluid phase ($a=0.051>a_c$, right).
In the solid phase, particles approximately return 
to their position after each cycle. In the fluid, 
there are regions of large displacements where irreversible rearrangements
take place.  The transition between reversible and irreversible 
phases at $a_c$ is discontinuous.
\label{fig:model}}
\end{figure}

We drive the system out of equilibrium by oscillating the diameter of each particle around its mean value $\sigma_i^0$, 
as shown in Fig.~\ref{fig:model}(a):
\begin{equation}
\sigma_i(t) = \sigma_i^0 \left[1 + a\cos\left( \omega t +
\psi_i\right)\right],
\label{eq:oscillation}
\end{equation} 
where $T = 2 \pi /\omega$ is the period of oscillation which we use as our time unit, 
and $a$ is an adimensional parameter which quantifies the 
intensity of the activity.
We impose very slow oscillations, $T \gg \tau_0$, 
such that the system is always located near an energy minimum
and inertial and hydrodynamic effects can be neglected. 
Specifically, we use $T = 820 \tau_0$. 
The average diameters $\sigma_i^0$ are drawn from a 
bidisperse distribution of diameters $0.71\sigma_0$ and $\sigma_0$ with 3:2 
proportion, in order to prevent crystallization. 
We use $\sigma_0$ as unit length.
We have introduced in Eq.~(\ref{eq:oscillation}) a random phase $\psi_i$ for each particle 
to constrain the total area fraction $\phi=\sum_{i} \frac{\pi\sigma_i^2(t)}{4L^2}$ to be strictly constant in time. 
The case with $\psi_i \equiv 0$ would correspond to affine compressions 
and expansions, 
which would then amount to studying the 
rheological response of the jammed solid forced at large scale, 
not an active material forced locally. 
We consider jammed systems with $\phi=0.94$, as appropriate for 
confluent tissues. Most simulations were 
performed with a very large system of $N=16000$ particles
(typically $L \approx 100\sigma_0$). We 
converged to this large value using simulations
with increasing sizes, seeking the disappearance of finite-size effects.
We also use finite-size scaling analysis to locate the 
phase transition with greater accuracy. 
For each $a$ value, we prepare fully random systems and apply 
the periodic perturbation until the system has reached steady state, 
either arrested or flowing. We then perform 
steady state measurements using averaging over time (in the flowing
phase), or over initial conditions (in the arrested phase). 

Figure~\ref{fig:model}(a) highlights the trajectory of a particle
during one period. We define the one-cycle displacement,
$\vec d_i(t) = \vec r_i(t+1) - \vec r_i(t)$,
as shown in Fig.~\ref{fig:model}(a). Collecting
the displacement of all particles we obtain the steady state one-cycle 
displacement map shown in Fig.~\ref{fig:model}(b) for both arrested 
and flowing phases. 
In the arrested phase, displacements are all very 
small and particles approximately return to the same 
position after each cycle, without undergoing configurational change.
On the other hand, at large activity  we observe regions of very large 
displacements where irreversible particle rearrangements
occur within one cycle. These local plastic events are spatially 
disordered, and they coexist with regions where displacements
are smaller: the dynamics is spatially heterogeneous.
Clearly, Fig.~\ref{fig:model} indicates the
existence of an arrested phase where particles do not move for small
$a$, and of a flowing phase for large $a$ 
where irreversible rearrangements take place during each cycle.
In the following we demonstrate  that the transition 
between these two regimes occurs at a well-defined 
activity value, $a_c$, and that it corresponds to a
first-order phase transition.

\begin{figure}
\begin{centering}
\includegraphics[width=8.5cm]{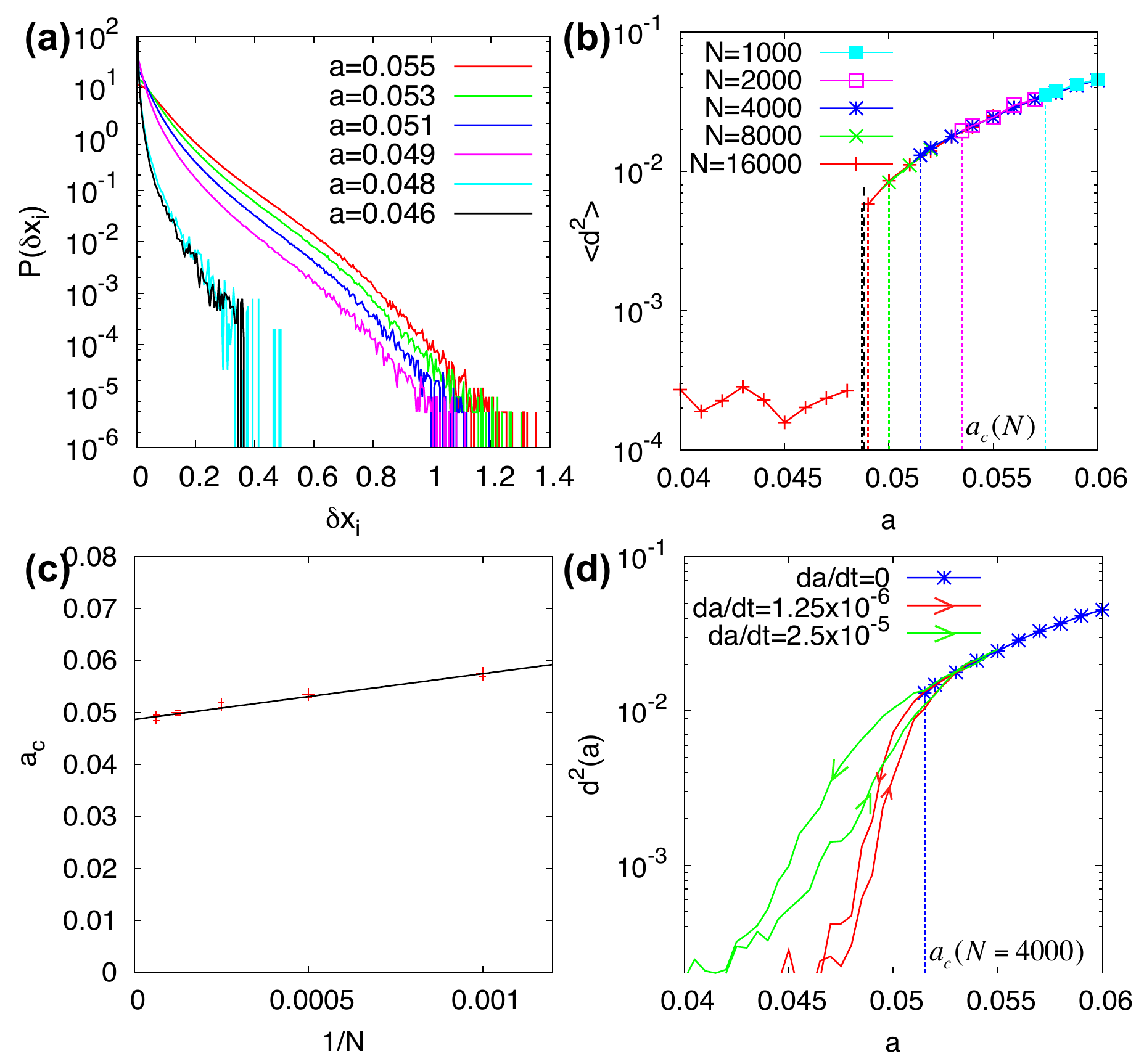}
\par\end{centering}
\protect\caption{
(a) Probability distribution of the one-cycle particle displacements $P(\delta x_{i})$ for different activities $a$.
The distribution changes discontinuously between flowing ($a>a_c \simeq 0.049$) and arrested ($a<a_c$) phases.
(b) Averaged one-cycle displacement squared at steady state for different 
system sizes $N$. 
$\langle d^2 \rangle$ is large above $a_c$, and  
drops discontinuously to nearly zero below $a_c$. 
The black vertical line represents $a_c(N \rightarrow \infty)$.
(c) The critical activity $a_c(N)$ tends to a 
finite value $\simeq 0.049$ as $1/N \rightarrow 0$.
(d) Evolution of $\langle d^2 \rangle$ 
for $N=4000$ as the activity is cycled 
at finite rate between $a=a_1>a_c$ and $a=a_2<a_c$.  
The blue curve represents the steady state value.
The hysteretic response gets sharper as $\frac{da}{dt}$ decreases. 
\label{fig:one-cycle}}
\end{figure}

We start by showing in Fig.~\ref{fig:one-cycle}(a) the probability 
distribution of one-cycle displacements, $P(\delta x_{i})$, 
where $\delta x_i = |x_i(t+1) - x_i(t)|$ (we use isotropy 
and average over $x$ and $y$ directions). 
In the flowing phase ($a>a_c \simeq 0.049$), $P(\delta x_i)$ has a
broad, nearly-exponential tail, stemming from 
particles involved in local rearrangements. As a result, all 
particles move significantly during each cycle. 
As shown below, the accumulation of these local plastic events 
over many cycles gives rise to diffusive behaviour and 
structural relaxation at large times. 
On the other hand in the solid phase ($a<a_c$), the exponential tail in 
$P(\delta x_{i})$ disappears and is replaced by a narrow Gaussian
distribution characterized by an average displacement per cycle 
that is considerably smaller. We show below that these small
displacements do not produce diffusive but only localised dynamics.

Crucially, the behaviour of $P(\delta x_i)$
changes abruptly when $a$ crosses $a_c$. 
We quantify this observation using a dynamic
order parameter for this phase change:
$\langle d^2 \rangle = \langle | \vec d_i(t) |^2 \rangle$,
where the brackets represent an average over time and particles
in steady state.
A similar quantity was defined in the context of the yielding 
transition in oscillatory shear~\cite{cipelletti,takeshi}.
In Fig.~\ref{fig:one-cycle}(b), we plot $\langle d^2 \rangle$ as a 
function of activity for different system sizes.
The flowing phase is characterized by large particle displacements
with $\left< d^2 \right>\gtrsim 0.01$,
whereas in the arrested phase, $\langle d^2 \rangle$ is about 100 times smaller.
Furthermore, $\left< d^2 \right>$ jumps discontinuously 
at a well-defined critical activity, $a_c(N)$.
To determine $a_c(N)$ we perform $8$ independent 
simulations from different initial configurations at each 
activity. We define $a_c$ such that the 8 runs 
remain diffusive for $a > a_c$ after a large time, $t=10^4$. 
In addition to the sharpness of the phase change, 
we also observe finite-size 
effects, since $a_c(N)$ decreases weakly with $N$.
As shown in Fig.~\ref{fig:one-cycle}(c), when plotted against
$1/N$, it is clear that $a_c$ extrapolates to a finite
value $a_c \approx 0.049$ when $N \to \infty$.

We substantiate further the discontinuous nature of the transition
by studying hysteresis effects. 
In Fig.~\ref{fig:one-cycle}(d), we measure how $\langle d^2 \rangle$ 
changes as we slowly cycle 
the activity $a(t)$ between $a=a_1>a_c$ and $a=a_2<a_c$ at 
a constant rate $\frac{da}{dt}$. 
We obtain the evolution for $\langle d^2 \rangle$ 
by averaging over $10^3$ such cycles.
We observe hysteresis cycles that become sharper
and narrower as the sweeping rate becomes slower. 
Such phenomenology is again representative 
of first-order phase transitions. 

\begin{figure}
\includegraphics[width=8.5cm]{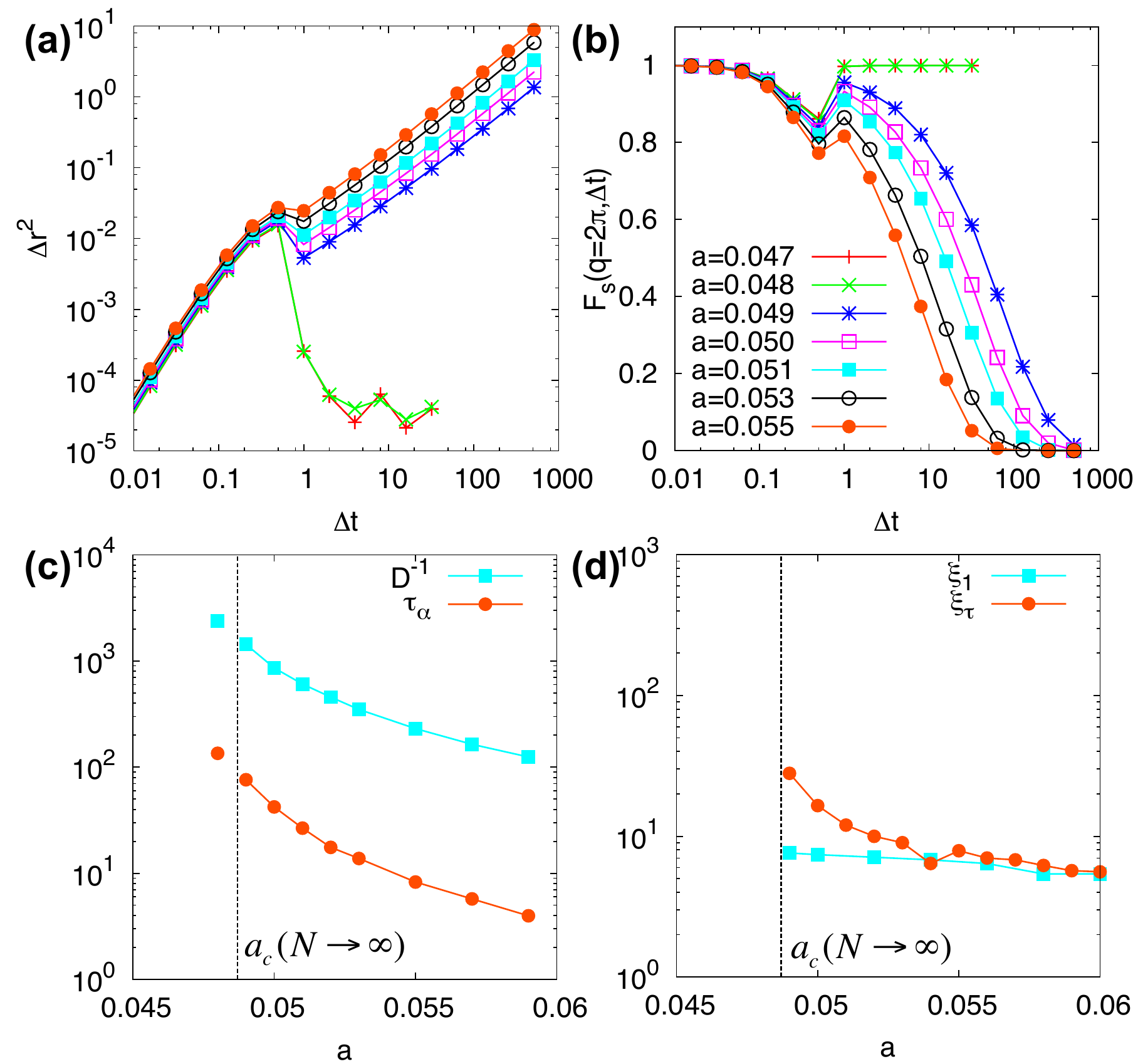}
\protect\caption{
(a) Mean-squared displacements 
are diffusive for $a>a_c$, but remain localised for $a<a_c$,
with a sharp discontinuity at $a_c$. 
(b) A similar discontinuous behaviour is observed for the 
self-intermediate scattering function $F_s(q,\Delta t)$,
which decays rapidly to 0 above $a_c$, but does not decay
below $a_c$. (c) Inverse diffusion constant $D^{-1}$ and relaxation 
time $\tau$ characterizing the long-time dynamics both 
increase modestly by about $1$ decade as $a \to a_c^+$. 
Below $a_c$ we find a metastable flowing phase where 
$D^{-1}$ and $\tau$ can be measured (isolated points)
before the system fully arrests. 
(d) Dynamic correlation lengthscales in the flowing phase 
measured over a time interval $\Delta t =1$ (for $\xi_1$) and
$\Delta t  = \tau$ (for $\xi_\tau$). Both lengths increase modestly 
towards $a_c$.
\label{fig:long-time}}
\end{figure}

We now turn to the long-time dynamics and  measure 
the mean-squared displacement (MSD) at steady state:
$\Delta r^2(\Delta t) =  \langle \left| 
\vec r_i(\Delta t)-\vec r_i(0)  \right| ^2 \rangle$,
see Fig.~\ref{fig:long-time}(a).
In the flowing phase ($a>a_c$), the system becomes diffusive at long times $\Delta r^2 (\Delta t \to \infty) \approx 4 D \Delta t$,
which defines the diffusion constant $D$. 
In the arrested phase instead, the MSD saturates to a small, finite value
at long times, demonstrating particle localisation in this regime.  
As a result, we find that $D>0$ above $a_c$ 
and $D=0$ below, with an abrupt change at $a_c$, as shown 
in Fig.~\ref{fig:long-time}(c). 
This sharp change in the long-time dynamics is also detected  
using the intermediate scattering function:
$F_s(q,\Delta t) = \left< e^{i \vec{q} \cdot \left[\vec 
r_i(\Delta t)-\vec r_i(0)  \right]} \right>$, which decays 
from 1 to 0 when particles move on average a distance $2\pi / |\vec{q}|$.
In Fig.~\ref{fig:long-time}(b), we show the time decay
of $F_s(q,\Delta t)$ for $q = 2\pi$, which corresponds 
to particles diffusing over a distance comparable to their diameter.
Above $a_c$, $F_s(q,\Delta t)$ decays to $0$ relatively rapidly.
We extract a relaxation time $\tau$ as $F_s(q,\tau) = 1/e$,
with again a discontinuous change between the two phases. 
We extract a relaxation time $\tau$ as $F_s(q,\tau) = 1/e$, and so
$\tau=\infty$ below $a_c$.

In Fig.~\ref{fig:long-time}(c) we report $D^{-1}$ and $\tau$ as a function of activity $a$.
Both measures of long-time dynamics increase modestly by about $1$ decade as  $a \to a_c^+$, and they do not diverge.
In addition, just below $a_c$, we find that the flowing phase can be `metastable' for a long time of order $30 \tau$ 
before suddenly evolving towards the arrested phase. 
Within this metastability window, long-time dynamical properties can be measured 
and we plot $D^{-1}$ and $\tau$ for this metastable liquid phase as 
isolated points in Fig.~\ref{fig:long-time}(c), which appear 
as the continuation of data at $a>a_c$. 
These observations confirm that both timescales do not diverge 
at $a_c$ and illustrate 
the first-order nature of the phase transition at $a_c$. 

Finally in Fig.~\ref{fig:long-time}(d) we plot two dynamic correlation lengthscales measured in the flowing phase 
which  only increase modestly without divergence at $a_c$. 
These dynamic lengthscales are obtained from analysis of a 4-point dynamic structure 
(see~\cite{berthier-book} for details about this classic measure of spatially heterogeneous dynamics).
In particular, we have studied dynamic correlations both over a delay time $\Delta t=1$ to probe 
spatial correlations of the one-cycle displacement map (see Fig.~\ref{fig:model}(b)), 
which gives us the one-cycle correlation length $\xi_{1}$. 
We also measured the dynamic lengthscale $\xi_\tau$ characterizing the long-time dynamics by setting $\Delta t=\tau$.
Both lengthscales increase modestly as $a \to a_c^+$, 
revealing collective motion in the flowing phase in the absence of  any criticality at the fluidisation transition. 

In conclusion, we have introduced a microscopic model for active materials where 
local energy injection stems from  active change of the particle sizes. 
This model is directly motivated by experimental observations of volume fluctuations of cells in epithelial tissues~\cite{zehnder,la-porta}. 
Our model exhibits a discontinuous non-equilibrium phase transition from an arrested to a fluid phase as 
the amplitude of self-deformation is increased. 
This active fluidisation transition is strikingly different from observations in self-propelled particles~\cite{RanNi,berthier-SP,szamel} 
in which continuous slowing down of several decades can be observed,
accompanied by growing dynamic lengthscales, 
reminiscent of glassy dynamics in dense fluids~\cite{berthier-nat-phys}. 
It also differs markedly from the static transition discovered in the vertex model, 
which is akin to a continuous rigidity transition~\cite{manning2}. 
We propose that a better analogy is with the yielding transition in soft amorphous solids, 
where irreversible rearrangements and particle diffusion result from applying a mechanical forcing above the yield stress. 
Evidence that yielding corresponds to a non-equilibrium dynamic first-order transition is mounting~\cite{cipelletti,takeshi,itamar}, 
the difference with our system being the scale at which the mechanical force is acting.  
Our model bears a strong resemblance to the fluid-like dynamics observed in biological tissues, 
where flow and diffusion also occur with finite correlation lengthscales and timescales~\cite{chavrier-tissues,silberzan,trepat-glassy-tissue}. 
We notice that the critical activity reported here is relatively small, $a_c \simeq 5\%$.
For comparison, certain eukaryotic cells have been observed to undergo volume fluctuations of up to $\sim20\%$~\cite{zehnder} 
due to the formation of various protrusions~\cite{la-porta}.
Such fluctuations could be large enough to fluidize dense suspensions of cells. 
Our results suggest that active volume fluctuations play an important role in the collective dynamics of epithelial tissues 
in addition to previously studied mechanisms, 
and provide a novel paradigm to interpret this dynamics in terms of an active yielding transition. 

\acknowledgments
We thank T. Kawasaki, M. E. Cates, and C. La Porta for discussion.
The research leading to these results has received
funding from the European Research Council under
the European Unions Seventh Framework Programme
(FP7/20072013)/ERC Grant Agreement No. 306845.


\begin{thebibliography}{99}

\bibitem{ramaswamy-review}
M. C. Marchetti, J. F. Joanny, S. Ramaswamy, T. B. Liverpool, J. Prost, 
M. Rao, and R. A. Simha,
Hydrodynamics of soft active matter, 
\emph{Rev. Mod. Phys.} \textbf{85} (2013) 1143.

\bibitem{vicsek-review}
T. Vicsek and A. Zafeiris, 
Collective motion,
{\it Phys. Rep.} {\bf 517} (2012) 71. 

\bibitem{chavrier-tissues}
M. Poujade, E. Grasland-Mongrain, A. Hertzog, J. Jouanneau, P. Chavrier, 
B. Ladoux, A. Buguin, and P. Silberzan,
Collective migration of an epithelial monolayer in response to a model wound,
\emph{Proc. Nat. Acad. Sci. (USA)} \textbf{104} (2007) 15988.

\bibitem{weibel}
M. F. Copeland and D. B. Weibel,
Bacterial swarming: a model system for studying dynamic self-assembly,
\emph{Soft Matter} \textbf{5} (2008) 1174.

\bibitem{ramaswamy-vibrated-rods}
V. Narayan, S. Ramaswamy, and N. Menon,
Long-lived giant number fluctuations in a swarming granular nematic,
\emph{Science} \textbf{317} (2007) 105.

\bibitem{olivier}
J. Deseigne, O. Dauchot, and H. Chat\'e,
Collective Motion of Vibrated Polar Disks
{\it Phys. Rev. Lett.} {\bf 105} (2010) 098001.

\bibitem{cecile}
I. Theurkauff, C. Cottin-Bizonne, J. Palacci, C. Ybert, and L. Bocquet,
Dynamic Clustering in Active Colloidal Suspensions with Chemical Signaling,
{\it Phys. Rev. Lett.} {\bf 108} (2012) 268303.

\bibitem{clemens}
I. Buttinoni, J. Bialke, F. Kümmel, H. Löwen, C. Bechinger, and T. Speck,
Dynamical clustering and phase separation in suspensions of self-propelled 
colloidal particles,
{\it Phys. Rev. Lett.}  {\bf 110} (2013) 238301.

\bibitem{silberzan}
L. Petitjean, M. Reffay, E. Grasland-Mongrain, M. Poujade, B. Ladoux, 
A. Buguin, P. Silberzan,  
Velocity Fields in a Collectively Migrating Epithelium,
{\it Biophys. Journal} {\bf 98} (2010) 1790.

\bibitem{mehes}
E. Mehes and T. Vicsek,
Collective motion of cells: from experiments to models,
{\it Integr. Biol.} {\bf 6} (2014) 831.



\bibitem{trepat-glassy-tissue}
T. E. Angelini, E. Hannezo, X. Trepat, M. Marquez, J. J. Fredberg, 
and D. A. Weitz,
Glass-like dynamics of collective cell migration,
\emph{Proc. Nat. Acad. Sci. (USA)} \textbf{108} (2011) 4714.

\bibitem{shraiman}
A. Puliafito, L. Hufnagel, P. Neveu, S. Streichan, A. Sigal, 
D. K. Fygenson, and B. I. Shraiman,
Collective and single cell behaviour in epithelial contact inhibition,
\emph{Proc. Nat. Acad. Sci. (USA)} \textbf{109} (2012) 739.

\bibitem{rorth}
P. Rorth,
Collective cell migration,
{\it Annu. Rev. Cell. Dev. Biol.} {\bf 25} (2009) 407.

% cell division
\bibitem{basan}
J. Ranft, M. Basan, J. Elgeti, J. -F. Joanny, J. Prost, F. Julicher,
Fluidization of tissues by cell division and apoptosis
{\it Proc. Nat. Acad. Sci. (USA)} {\bf 107} (2010) 20863.

% cell crawling
\bibitem{theriot}
S. M. Rafelski and J. A. Theriot,
Crawling toward a unified model of cell motility:
Spatial and temporal regulation of actin dynamics,
{\it Annu. Rev. Biochem.} {\bf 73} (2004) 209.

% volume fluctuations
\bibitem{zehnder}
S. M. Zehnder, M. Suaris, M. M. Bellaire, and T. E. Angelini,
Cell Volume Fluctuations in MDCK Monolayers,
{\it Biophys. Journal} {\bf 108} (2015) 247.

\bibitem{la-porta}
A. Taloni, E. Kardash, O. U. Salman, L. Truskinovsky, S. Zapperi, 
C. A. M. La Porta,
Volume Changes During Active Shape Fluctuations in Cells,
\emph{Phys. Rev. Lett.} \textbf{114} (2015) 208101.

\bibitem{yamamoto}
S. K. Schnyder, Y. Tanaka, J. J. Molina, and R. Yamamoto,
Collective motion of cells crawling on a substrate: roles of cell shape 
and contact inhibition,
\emph{arXiv:1606:07618}.

\bibitem{manning}
D. Bi, J. Lopez, J. Schwartz, and M. L. Manning,
Energy barriers and cell migration in densely packed tissues,
\emph{Soft Matter} \textbf{10} (2014) 1885.

\bibitem{julicher}
R. Farhadifar, J.-C. Roper, B. Aiguoy, S. Eaton, and F. J\"ulicher,
The Influence of Cell Mechanics, Cell-Cell Interactions, and Proliferation 
on Epithelial Packing
\emph{Curr. Biol.} \textbf{17} (2007) 2095.

\bibitem{manning2}
D. Bi, J. H. Lopez, J. M. Schwarz, and M. L. Manning,
A density-independent rigidity transition in biological tissues,
{\it Nature Phys.} {\bf 11} (2015) 1074.

\bibitem{vicsek}
T. Vicsek, A. Czirok, E. Ben-Jacob, I. Cohen, and O. Shochet,
Novel type of phase transition in a system of self-driven particles,
\emph{Phys. Rev. Lett.} \textbf{75} (1995) 1226.

\bibitem{marchetti-active-jamming}
S. Henkes, Y. Fily, and M. C. Marchetti,
Active jamming: Self-propelled soft particles at high density,
\emph{Phys. Rev. E} \textbf{84} (2011) 040301.

\bibitem{berthier-nat-phys}
L. Berthier and J. Kurchan,
Non-equilibrium glass transitions in driven and active matter,
\emph{Nature Phys.} \textbf{9} (2013) 310.

\bibitem{RanNi}
R. Ni, M. A. Cohen Stuart, and M. Dijkstra,
Pushing the glass transition towards random close packing using 
self-propelled hard spheres, 
\emph{Nature Comm.} \textbf{4} (2013) 2704.

\bibitem{berthier-SP}
L. Berthier,
Nonequilibrium glassy dynamics of self-propelled hard disks, 
\emph{Phys. Rev. Lett.} \textbf{112} (2014) 220602.

\bibitem{szamel}
E. Flenner, G. Szamel, and L. Berthier,
The nonequilibrium glassy dynamics of self-propelled particles,
{\it arXiv:1606.00641}.

\bibitem{vedula}
S. R. Vedula, M. C. Leong, T. L. Lai, P. Hersen, A. J. Kabla, 
C. T. Lim and B. Ladoux,
Emerging modes of collective cell migration 
induced by geometrical constraints,
\emph{Proc. Nat. Acad. Sci. (USA)} {\bf 109} (2012) 12974.

\bibitem{garcia}
S. Garcia, E. Hannezo, J. Elgeti, J.-F. Joanny, P. Silberzan, 
and N. S. Gov,
Physics of active jamming during collective cellular motion in a monolayer,
\emph{Proc. Nat. Acad. Sci. (USA)} {\bf 112} (2015) 15314.

\bibitem{sepulveda}
N. Sepulveda, L. Petitjean, O. Cochet, E. Grasland-Mongrain, P. Silberzan,
and V. Hakim, 
Collective Cell Motion in an Epithelial Sheet Can Be Quantitatively 
Described by a Stochastic Interacting Particle Model,
{\it PLoS Comput. Biol.} {\bf 9} (2013) e1002944.


\bibitem{bonn}
D. Bonn, J. Paredes, M. M. Denn, L. Berthier, T. Divoux, and S. Manneville,
Yield Stress Materials in Soft Condensed Matter,
{\it arXiv:1502.05281}.

\bibitem{durian}
D. J. Durian,
Foam mechanics at the bubble scale, 
\emph{Phys. Rev. Lett.} \textbf{75} (1995) 26.

\bibitem{cipelletti}
E. D. Knowlton, D. J. Pine, and L. Cipelletti,
A microscopic view of the yielding transition in concentrated emulsions,
\emph{Soft Matter} \textbf{10} (2014) 6931.

\bibitem{takeshi}
T. Kawasaki and L. Berthier,
Macroscopic yielding in jammed solids is accompanied by a 
non-equilibrium first-order transition in particle trajectories,
{\it arXiv:1507.04120}.

\bibitem{berthier-book}
\emph{Dynamical heterogeneities in glasses, colloids and granular materials},
Eds: L. Berthier, G. Biroli, J.-P. Bouchaud, L. Cipelletti, and W. van Saarloos 
(Oxford University Press, Oxford, 2011). 

\bibitem{itamar}
P. Jaiswal, I. Procaccia, C. Rainone, and M. Singh, 
Mechanical Yield in Amorphous Solids: 
a First-Order Phase Transition, 
{\it Phys. Rev. Lett.} {\bf 116} (2016) 085501.




\end{thebibliography}
\end{document}